\documentclass{revtex4}
\usepackage{graphics,amssymb}
\newcommand{\Zeta}{Z}
\begin{document}
\title{An experimental study to discriminate between the validity of diffraction theories for off-Bragg replay\thanks{In honour of Romano A. Rupp on the occasion of his 60$^{th}$ birthday.}}
\author{Martin Fally}
\affiliation{University of Vienna, Faculty of Physics, Physics of Functional Materials, Boltzmanngasse 5, A-1090 Wien, Austria}
\homepage{http://fun.univie.ac.at}
\email{martin.fally@univie.ac.at}
\author{J\"urgen Klepp}
\affiliation{University of Vienna, Faculty of Physics, Physics of Functional Materials, Boltzmanngasse 5, A-1090 Wien, Austria}
\email{Juergen.Klepp@univie.ac.at}
\author{Yasuo Tomita}
\affiliation{Department of Engineering Science, University of Electro-Communications, 1-5-1 Chofugaoka, Chofu, Tokyo 182, Japan}
\email{ytomita@ee.uec.ac.jp}
%
%
%
%
%

%
\begin{abstract}
We show that experiments clearly verify the assumptions made by the first-order two-wave coupling theory for one dimensional lossless unslanted planar volume holographic gratings using the beta-value method rather than Kogelnik's K-vector closure method. Apart from the fact that the diffraction process is elastic, a much more  striking difference between the theories becomes apparent particularly in the direction of the diffracted beam  in off-Bragg replay.  We therefore monitored the direction of the dif\-frac\-ted beam as a function of the off-Bragg replay angle in two distinct cases: \ref{it:1} the diffracted beam lies in the plane of incidence and \ref{it:2} the sample surface normal, the grating vector and the incoming beam do not form a plane which calls for the vectorial theory and results in conical scattering.
\end{abstract}
\maketitle
\section{Introduction\label{intro}}
Theories to analyze diffraction by sinusoidal structures making use of coupled waves have been around since the thirties for diffraction of light by standing sound waves \cite{Raman-piasa35,Phariseau-piasa65}. In the late sixties experimental holography became popular with the invention of lasers leading to a resurgence of coupled wave theories to describe the diffraction of light by plane-wave holograms \cite{Kogelnik-Bell69}. Interestingly, even much earlier an alternative approach using coupled modes (dynamical theory of diffraction, DDT) was favored for  X-rays  \cite{Ewald-adp16.1}, which was applied later also for neutrons  \cite{Rauch78}.
 Though {\em Moharam and Gaylord} offer an exact solution to the diffraction problem - the rigorous coupled wave analysis (RCWA) - \cite{Moharam-josa81}, and demonstrated that the DDT is completely equivalent in its rigorous variant  \cite{Gaylord-apb82}, still approximate theories are mostly used for evaluation of experimental data. Among those the first-order two-wave coupling theory of {\em Kogelnik} turned to the favorite one in holography \cite{Kogelnik-Bell69}. Here, the second-order derivatives of the amplitudes are neglected on the basis of the slowly varying envelope approximation. Its  secret of success seems to be the simplicity of the relevant equations and the comprehensive treatment of a number of useful cases (transmission-reflection gratings, slanted gratings, phase- and absorption gratings) for in-Bragg and off-Bragg replay. Despite the indisputable merits, soon a variant of the theory was published  \cite{Uchida-josa73} that replaces one assumption made by {\em Kogelnik} for off-Bragg 
replay -  the K-vector closure method (KVCM) - by the beta-value method (BVM).  It was clearly pointed out by {\em Syms} that from a mathematical point of view a class of valid first-order theories exists to which both of them belong \cite{Syms-90}.
  The fundamental difference arises from neglecting (KVCM) or properly taking into account (BVM) the boundary conditions. A comparison of these  approximate first-order theories with the RCWA led to the conclusion 'that the BVM offers definite advantages over the KVCM method' \cite{Sheridan-jmo92}.  The BVM later was also strongly promoted by the books of {\em Solymar}  \cite{Solymar-02} and {\em Yeh} \cite{Yeh-93} as well as by setting up a first-order two-wave coupling theory for anisotropic media  \cite{Montemezzani-pre97}. Any of these above mentioned publications focuses on the calculation of the amplitudes for the diffracted waves. Both methods give similar results for the amplitudes, but strongly disagree when it comes to the directions of the diffracted beam for off-Bragg replay.

 In what follows we will demonstrate that experiments focussing on the direction of the first-order diffracted beam falsify the KVCM, as presumed previously in the case of X-rays  \cite{Sheppard-ije76}, and show that the direction of the diffracted beam follows the predictions of the BVM. We start with a summary of the predictions given by the two theories. Next, the results of diffraction experiments from an unslanted volume transmission grating and a comparison of the experiments with the KVCM and BVM are performed. We finish with a discussion and conclusion.
\section{First-order two-wave coupling diffraction theories: summary \label{sec:2}}
The emphasis of this part lies on working out the difference between the two theories for the direction of the diffracted beam. The issues related to the differences on the amplitudes and phases were already addressed in detail theoretically and experimentally , see e.g. Refs.  \cite{Sheridan-jmo92,Syms-ao83}.  The  conclusion was that they do not differ when obeying the Bragg condition, however, the BVM having some advantages over the KVCM for a hologram in off-Bragg replay.
The first-order diffraction efficiency for an unslanted phase grating in transmission is
\begin{equation}
\label{eq:diffeff}
  \eta_{\pm1}=\frac{c_R}{c_S}\nu^2 {\rm sinc}^2{\sqrt{\nu^2+\xi^2}}.
\end{equation}
with $\nu=n_1\pi d/\lambda \sqrt{c_R/c_S}, c_R=\cos{\theta}$. Here, $n_1$ is the refractive-index modulation, $d$ the grating thickness, $\lambda$ the wavelength in free space, and $\theta$ the angle of incidence in the medium.
The parameters $c_S,\xi$ for each theory  are defined below in the corresponding sections \ref{sec:KVCM} and \ref{sec:BVM}, respectively.
We show wavevector diagrams yielding the expected direction of the diffracted beam for the employed experimental configurations and a simple analytic formula for the (external) excess angle $\varepsilon$, i.e., the angle exceeding the value of twice the Bragg angle between the zero order and first diffraction orders. The notation is as follows: wavevectors {\em inside the medium} are denoted by $\vec{q}$, {\em in free space} by $\vec{k}$; angles inside are $\theta$ and $\theta_B$; angles outside are $\Theta_B,\Theta,\Delta\Theta$ and $\Zeta$, to be defined in the next sections.
                    \subsection{Experimental geometries\label{sec:expgeo}}
\begin{figure}[p]
\resizebox{\columnwidth}{!}{%
\includegraphics{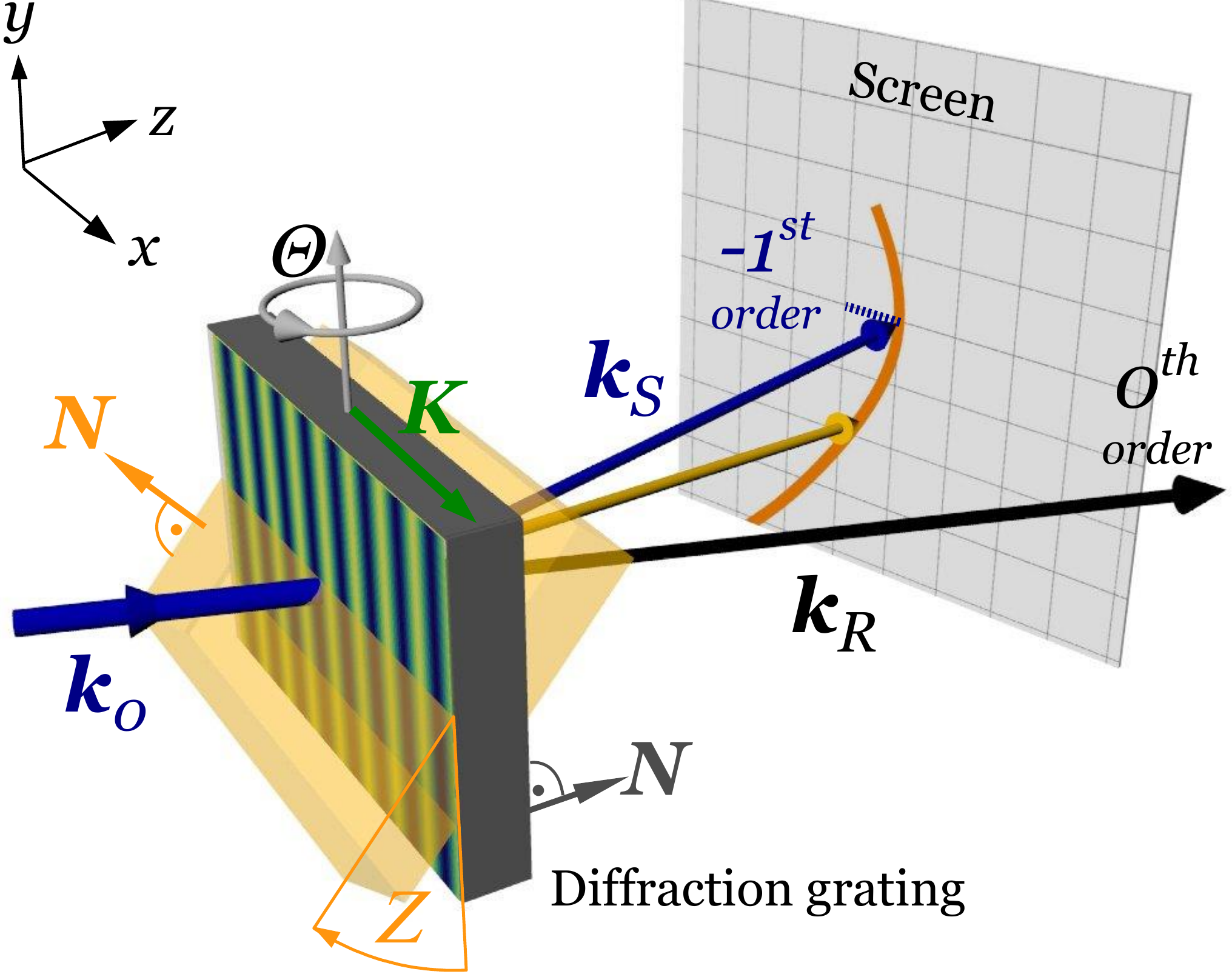}
}
\caption{Schematic of the setup in an off-Bragg replay diffraction experiment for planar and tilt geometry. For the notation confer to the text. }
\label{fig:geo}       
\end{figure}

Experiments were realized for two different geometries that are sketched in Fig. \ref{fig:geo}:
\renewcommand{\theenumi}{[\alph{enumi}]}
\begin{enumerate}
\item Planar geometry: the wavevector $\vec{k}_0$ of the incident beam, grating vector $\vec{K}$ and sample surface normal $\vec{N}$ are coplanar in the $x-z$-plane\label{it:1}
\item Tilt geometry: $\vec{k}_0$ and $\vec{K}$ lie in the plane $x-z$ whereas $\vec{N}$ is rotated by an angle $\Zeta$ around the grating vector. This is equivalent to the experimental situation depicted in Fig. 3 of Ref.  \cite{Syms-oa85} and treated in a rigorous manner in Ref.  \cite{Moharam-josa83.2}.\label{it:2}
\end{enumerate}
Note that for an unslanted grating $\vec{K}\cdot\vec{N}\equiv 0$ applies.
For the discussion that follows we assume the sample to be fixed in the reference frame while the incident beam rotates in the $x-z$ plane in contrast to the experimental situation shown in Fig. \ref{fig:geo} where instead the sample is rotated while keeping the incident beam fixed.
        \subsection{KVCM\label{sec:KVCM}}
Let us define the parameters for Eq. (\ref{eq:diffeff}) as $c_S=c_R$ and $\xi= (\theta-\theta_B)|\vec{K}| d/2$.
\begin{figure}[b]
\centering
\resizebox{10cm}{!}{%
\fbox{\includegraphics{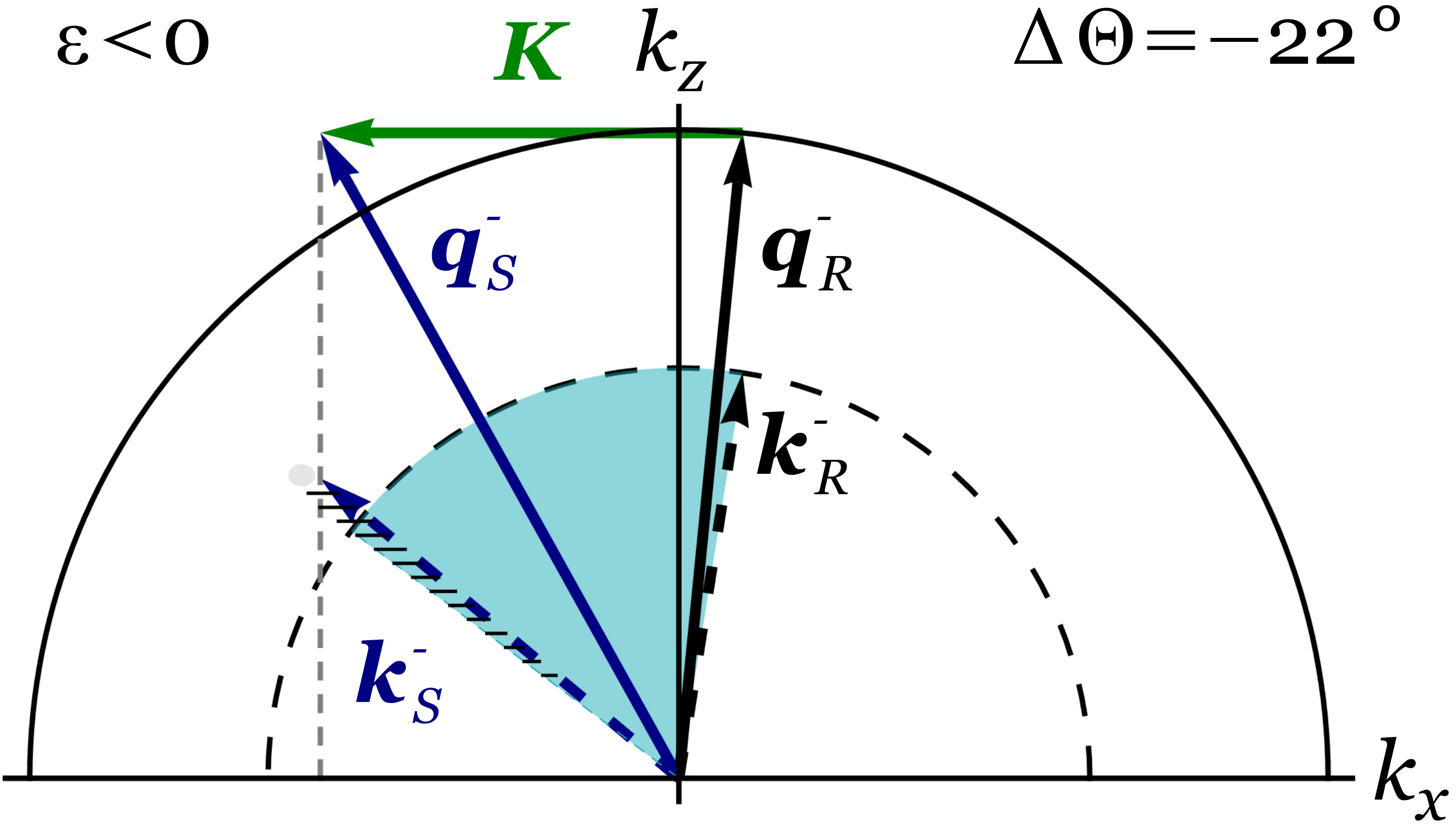}}
}
\resizebox{10cm}{!}{%
\fbox{\includegraphics{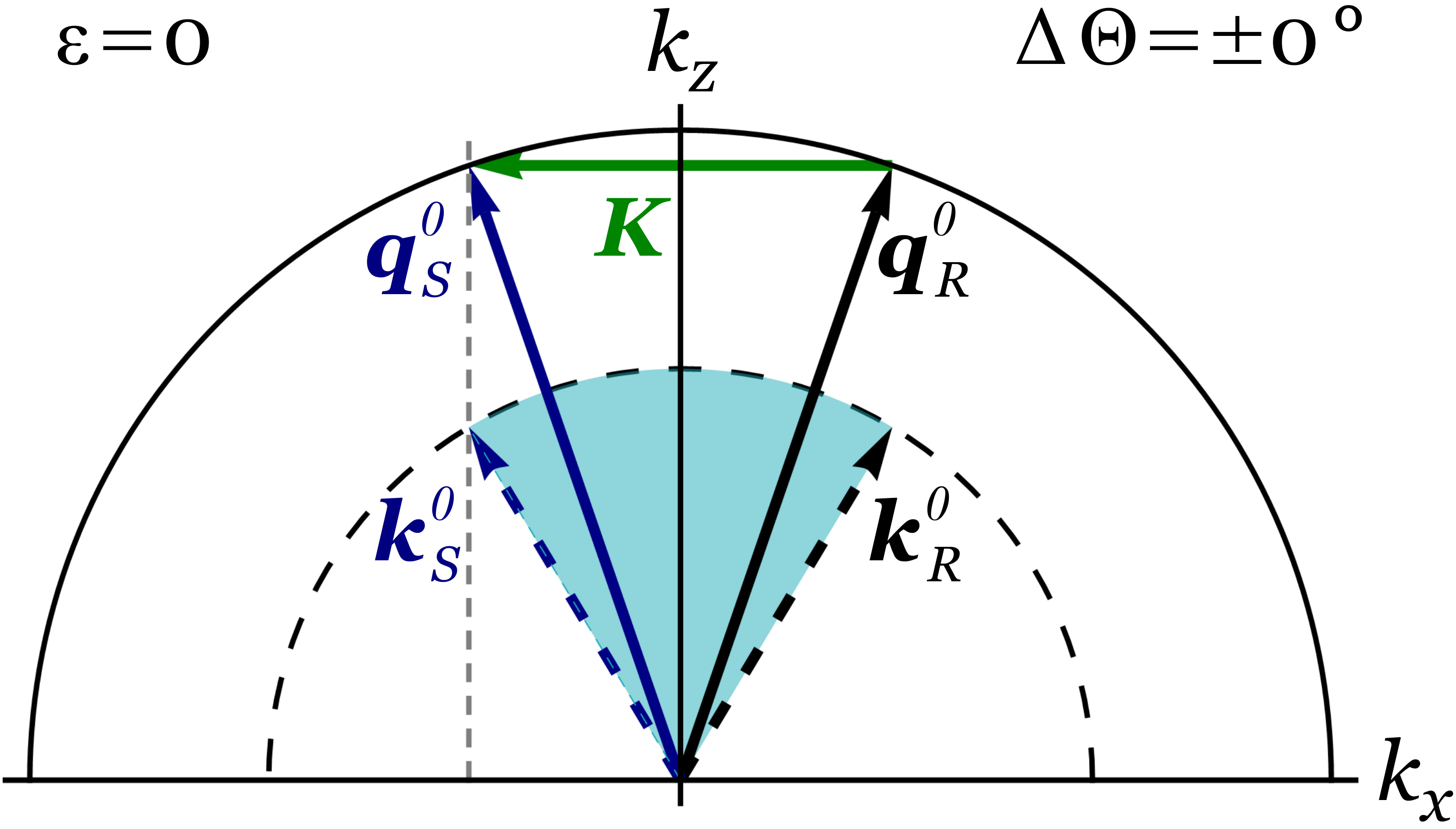}}
}
\resizebox{10cm}{!}{%
\fbox{\includegraphics{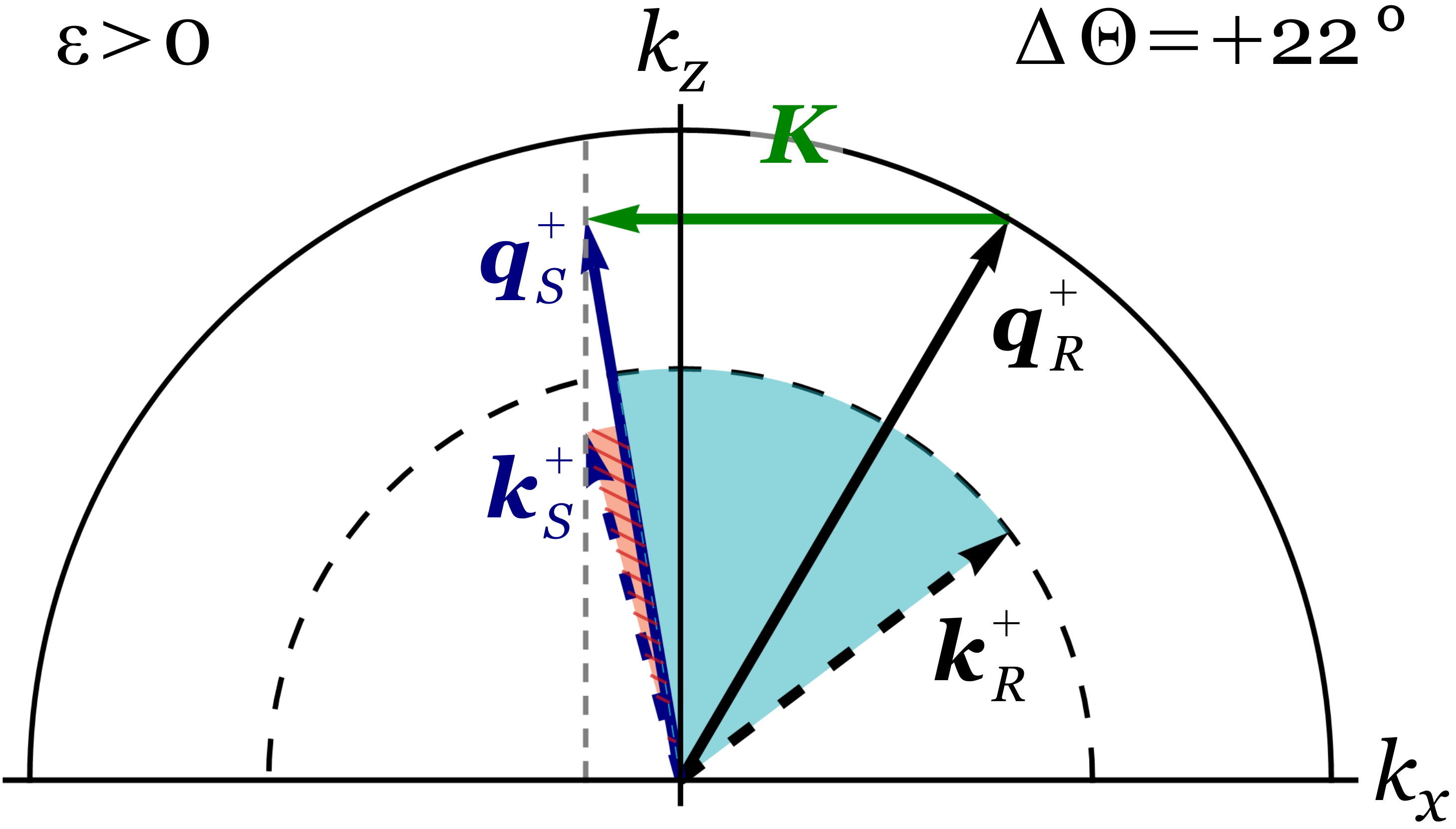}}
}
\caption{KVCM: wavevector diagram for three angles of incidence: $\Delta\Theta=\Theta-\Theta_B=-22,0,+22^\circ$. The full (dashed) arrows represent the wavevectors inside (outside) the medium. The blue shaded sector covers $2\Theta_B$, whereas the hatched sector gives the 'excess' angle $\varepsilon$. Note, that $\varepsilon<0$ for $\Delta\Theta<0$.}
\label{fig:kvc}       
\end{figure}
For the wavevector $\vec{q}_S$ of the diffracted beam the KVCM simply assumes
\begin{equation}\label{eq:qsKVCM}
  \vec{q}_S=\vec{q}_R\pm\vec{K},
\end{equation}
with $\vec{q}_R$ the wavevector of the forward diffracted beam.
The corresponding wavevector diagrams for the diffraction process are shown in Fig. \ref{fig:kvc}.
As pointed out earlier  \cite{Syms-ao83}, Eq. (\ref{eq:qsKVCM}) comprises a valid choice to obtain a 1-D scalar wave equation. However, taking Eq. (\ref{eq:qsKVCM})  seriously, the diffraction would be 'inelastic' for off-Bragg replay \cite{Sheridan-jmo92}, i.e., the wavelength of the diffracted beam is expected to change upon rotation.
For the parameters of the experiment  described in section \ref{sec:exp} the wavelength change at $\pm 20^\circ$ off-Bragg angle would be about $\pm$10\%, so that the incident green beam color should change to either deep blue or yellow, respectively.
The expected 'excess' angle reads:
\begin{eqnarray}\label{eq:excessKVCM}
\varepsilon(\Theta)&=&\arcsin{\left[
\frac{n (2 \sin{\Theta_B}-\sin{\Theta})}{\sqrt{n^2-\sin^2{\Theta}+(2 \sin{\Theta_B}-\sin{\Theta})^2}}
\right]}
+\Theta-2\Theta_B,
\end{eqnarray}
                        \subsection{BVM\label{sec:BVM}}
For the BVM the parameters for Eq. (\ref{eq:diffeff}) are $\xi=(\cos{\theta}-c_S)\beta d/2$ and  $c_S=\sqrt{1 - (2 \sin{\theta_B - \sin{\theta}})^2}$ with $\beta=2\pi n/\lambda$. The latter parameter, which is eponymous to the method, is the modulus of the wavevectors {\em in the medium} with average refractive index $n$ and $\sin{\theta_B}=\pm\lambda|\vec{K}|/(4\pi n)$.
\begin{figure}\centering
\resizebox{10cm}{!}{%
\fbox{\includegraphics{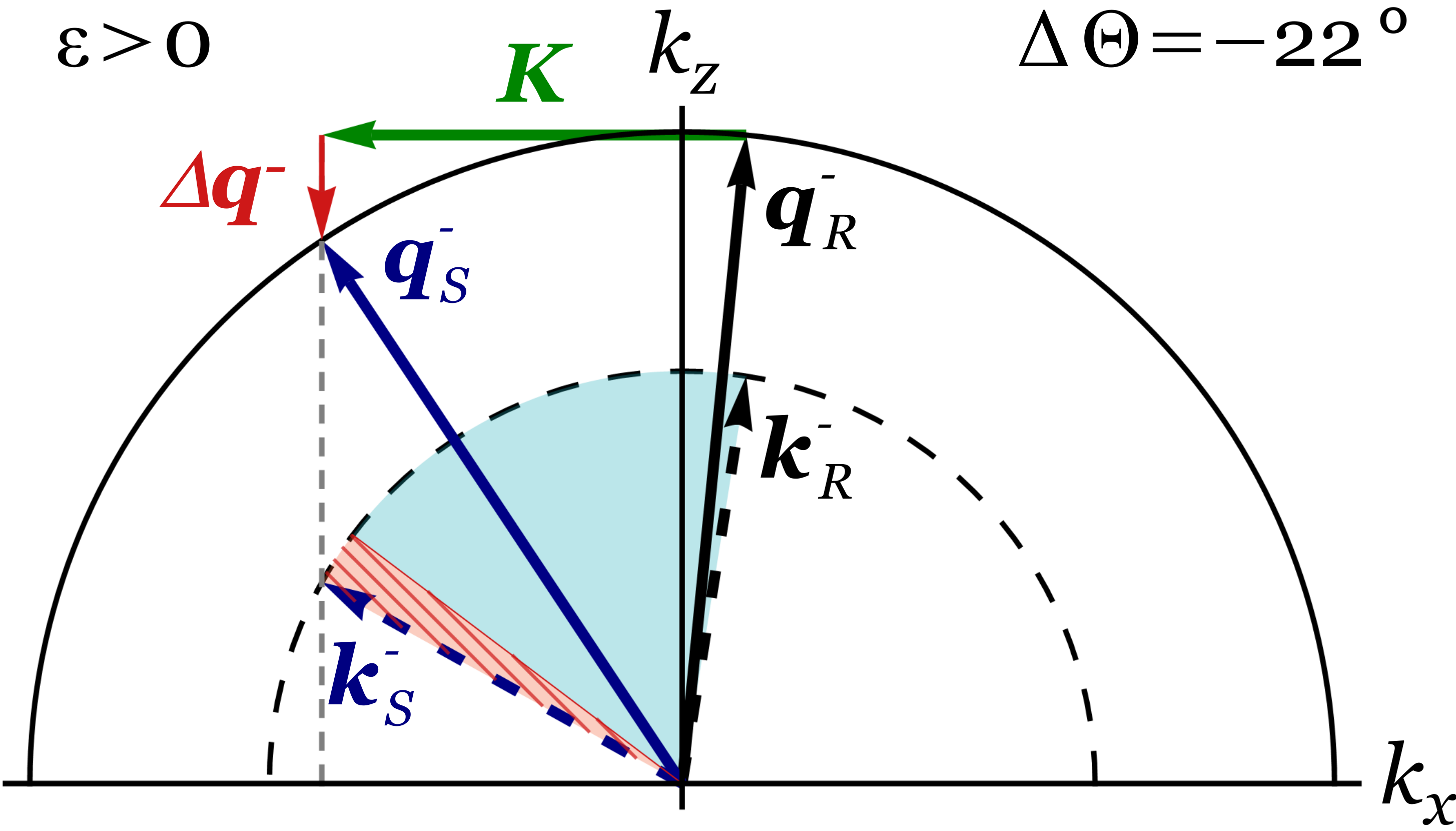}}
}
\resizebox{10cm}{!}{%
\fbox{\includegraphics{wv_bvm-0_cropped.pdf}}
}
\resizebox{10cm}{!}{%
\fbox{\includegraphics{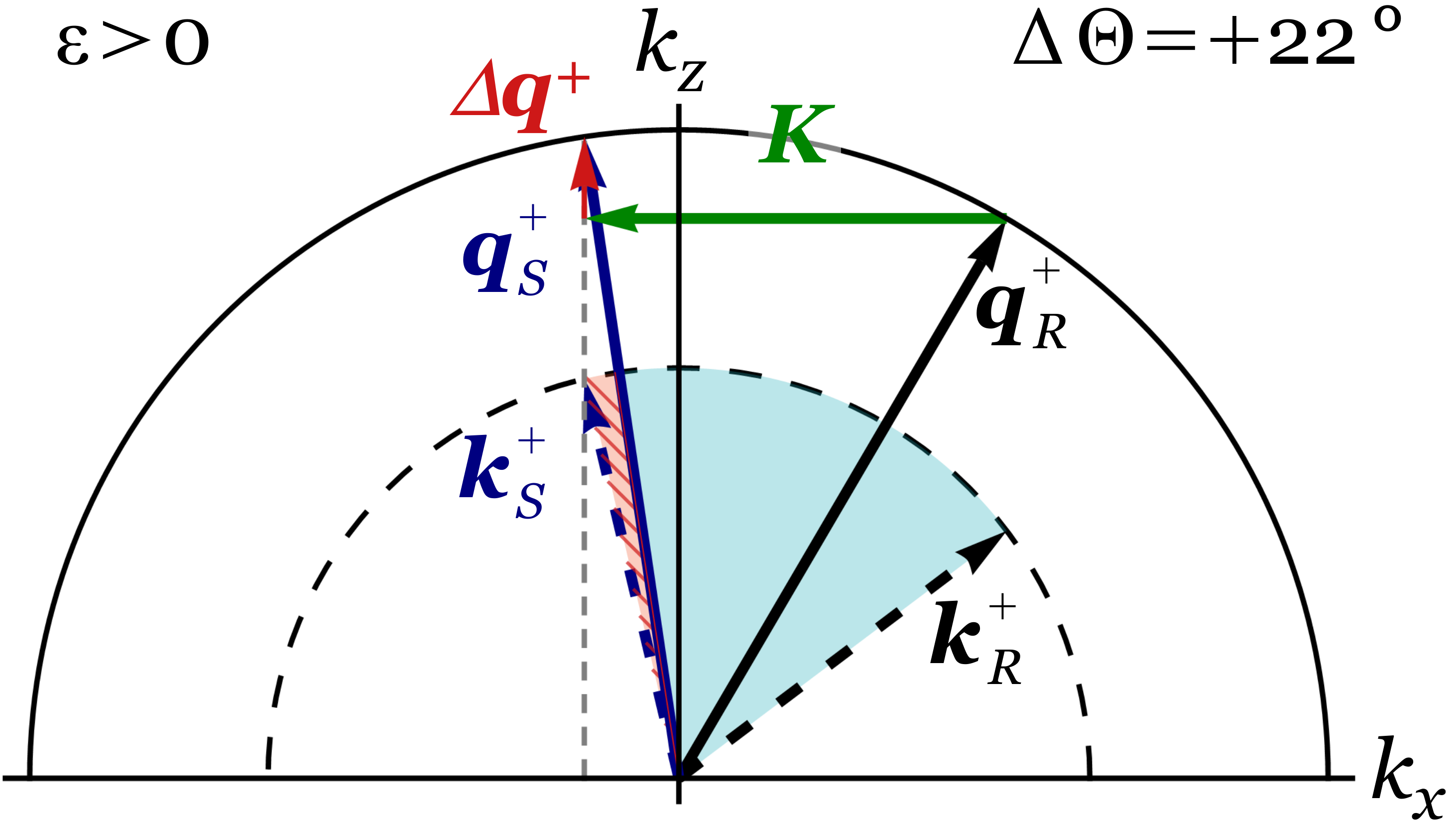}}
}
\caption{BVM: wavevector diagram for three angles of incidence: $\Delta\Theta=\Theta-\Theta_B=-22,0,+22^\circ$. The full (dashed) arrows represent the wavevectors inside (outside) the medium, respectively.
The blue shaded sector covers $2\Theta_B$, whereas the hatched sector represents the excess angle $\varepsilon> 0$.}
\label{fig:2dbvm}       
\end{figure}
The Floquet-condition in the BVM case requires the wavevector of the diffracted beam to obey
\begin{equation}\label{Eq:qsBVM}
  \vec{q}_S=\vec{q}_R\pm\vec{K}\pm\Delta q\vec{N}
\end{equation}
with
\begin{equation}
\Delta q(\theta)=\beta \left(\cos{\theta} -\sqrt{1 - (2 \sin{\theta_B - \sin{\theta}})^2}\right).
\end{equation}
At first let us consider the planar geometry \ref{it:1} as defined in Sec. \ref{sec:expgeo}. By geometric reasoning and applying Snell's law at the boundaries we obtain the (external)  excess angle $\varepsilon$ as a function of the (external) off-Bragg angle $\Delta\Theta$ as
\begin{equation}\label{eq:excess2d}
  \varepsilon(\Theta)=\arcsin{\left[2 \sin{\Theta_B}-\sin{\Theta}\right]}+\Theta-2\Theta_B.
\end{equation}
We want to draw the attention to the fact that $\varepsilon$ is independent of the refractive index of the medium because of phase-matching at the boundaries. The wavevector diagrams for $\Delta\Theta=-22,0,+22^\circ$ are shown in Fig. \ref{fig:2dbvm}.
It can be seen that for $q_{S,x}>2\pi/\lambda$ the diffracted beam is excited but totally reflected in the medium.
The corresponding rotation angle for which total reflection occurs is
\begin{equation}\label{eq:cutoff2D}
  \Theta_{TR}=\arcsin{(2 \sin{\Theta_B}\mp 1)}-\Theta_B.
\end{equation}
Now, let us turn to the more general case, i.e., configuration \ref{it:2} defined in Sec. \ref{sec:expgeo}. By inspecting Fig. \ref{fig:3dbvm} it is obvious that now the diffracted beam moves out of the $x-z$ plane upon off-Bragg replay.
\begin{figure}
\resizebox{\columnwidth}{!}{%
\includegraphics{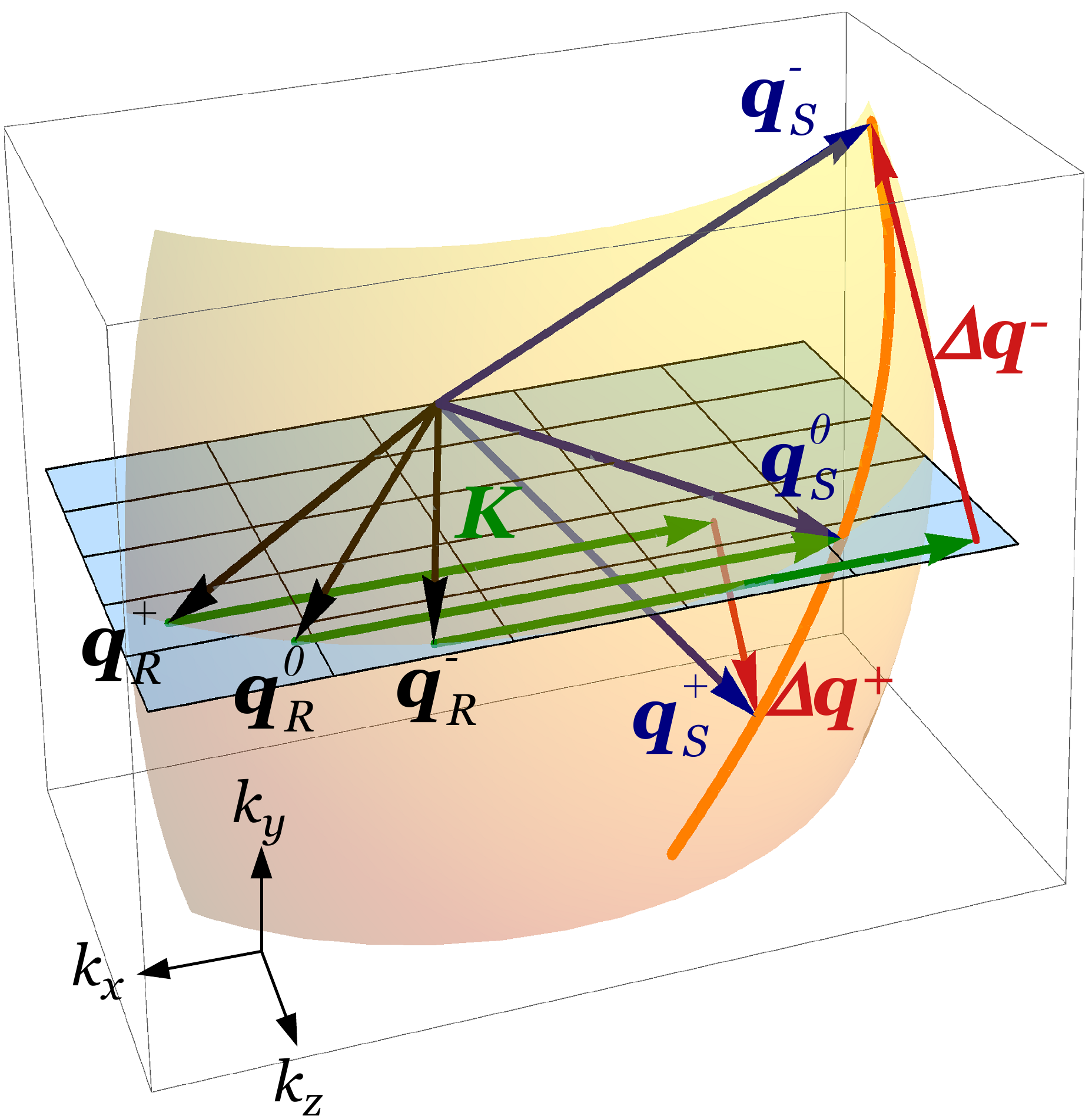}
}
\caption{Wavevector diagram for the tilt geometric configuration \ref{it:2}  for three rotation angles ($\Delta\Theta=0,\pm 17^\circ$) where $\vec{q}^{0,\pm}$ denote the corresponding wavevectors. The orange curve gives the tie-points of the diffracted wavevectors $\vec{q}_S$ for off-Bragg replay. Tilt angle $\Zeta=45^\circ$.}
\label{fig:3dbvm}       
\end{figure}
The direction of the diffracted beam can be analytically expressed as a function of the two external angles $\Theta$ and $\Zeta$ characterizing the incident beam. The three-step-procedure to arrive at this is as follows: define the external wavevector $\vec{k}_0$ and the plane of incidence in terms of $\Theta$ and $\Zeta$, apply Snell's law at the entrance boundary and make use of Eq. (\ref{Eq:qsBVM}) to derive the internal wavevectors $\vec{q}_{R,S}$. Finally we employ Snell's law once more for the diffracted beams to end up with $\vec{q}_{R,S}$. Again - as expected - the direction of the diffracted beams does not depend on the refractive index of the medium. The result is lengthy and therefore we do not give it explicitly here.
We also find an angle of total reflection $\Theta_{TR}$ that depends on the tilting angle $\Zeta$ or vice versa a cutoff $\Zeta_{TR}(\Theta)$
\begin{equation}\label{eq:cutoffgen}
\Zeta_{TR}(\Theta)=\arccos{\left[\frac{2\sqrt{\sin{\Theta_B} (\sin{\Theta_B}-\sin{\Theta})}}{\cos{\Theta}}\right]}.
\end{equation}
        \subsection{Comparison\label{sec:compare}}
\begin{figure}[b]
\resizebox{\columnwidth}{!}{%
\includegraphics{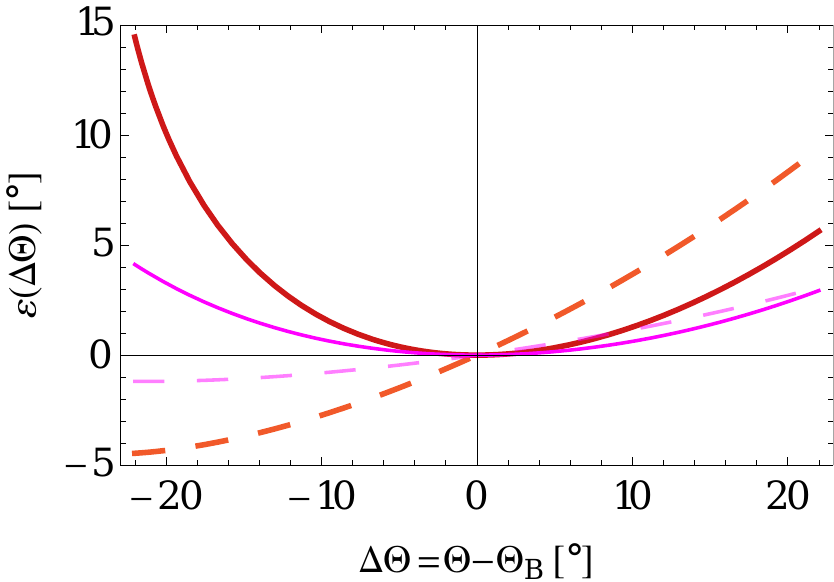}
}
\caption{Excess angle $\varepsilon(\Delta\Theta)$ for the BVM (Eq. \ref{eq:excess2d}, solid lines) and the KVCM (Eq. \ref{eq:excessKVCM}, dashed lines) for two wavelengths: 351 nm (thin)  and 633 nm (thick).}
\label{fig:2ddiff}       
\end{figure}

\begin{figure}[b]
\resizebox{0.575\columnwidth}{!}{%
\includegraphics{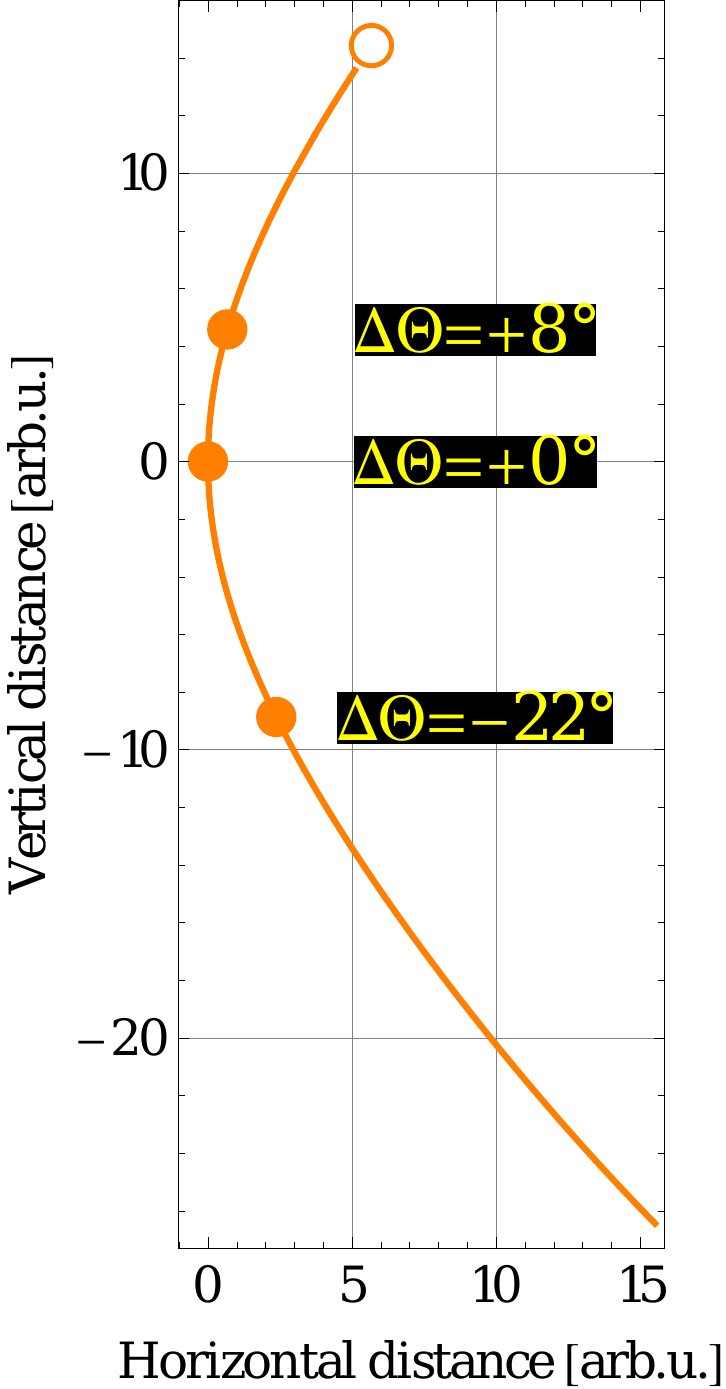}
}
\resizebox{0.35\columnwidth}{!}{%
\fbox{\includegraphics{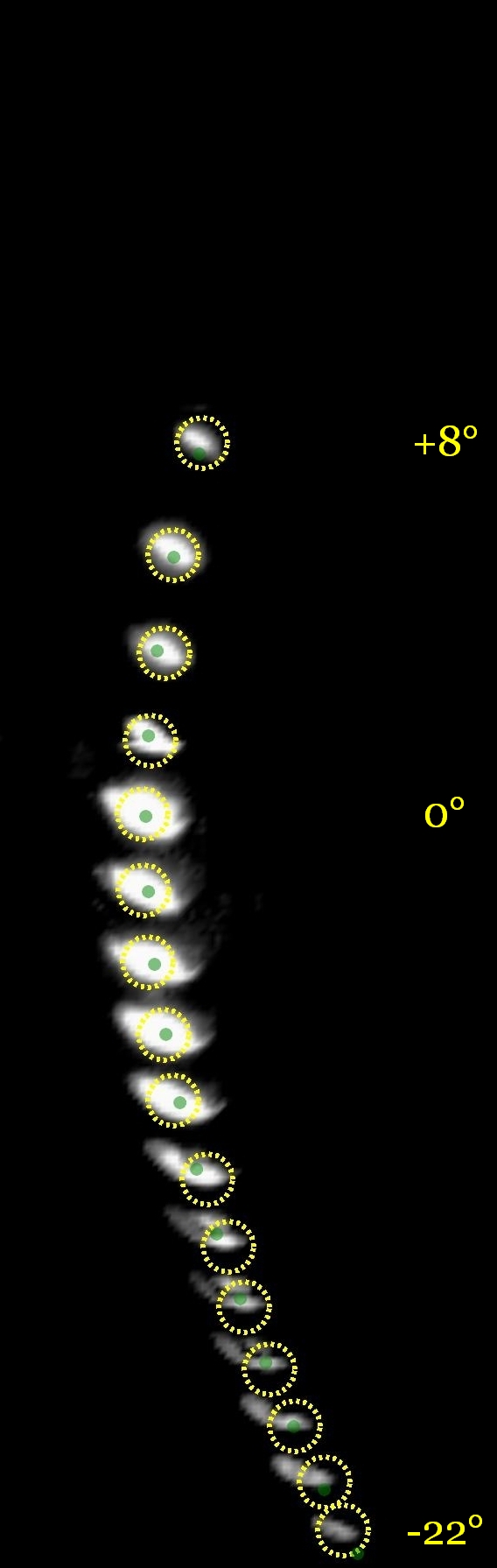}}
}
\caption{{\it Left}: Diffracted beam on a screen for off-Bragg replay according to BVM in configuration \ref{it:2}. The screen is perpendicular to the diffracted beam for in-Bragg.{\it Right}: Measured position of the diffracted beams on a screen for a sequence of rotation angles $\Theta=-22\ldots+8$ with an increment of 2$^\circ$ and a tilt angle of $\Zeta=41.7^\circ$. Dashed circles are the centers of gravity of the spots, the solid disks are the fitted results.}
\label{fig:3donscreen}       
\end{figure}

For the planar geometry both theories correctly predict that the direction of the diffracted beam lies in the plane of incidence, i.e., $\vec{k}_R,\vec{N},\vec{K},\vec{k}_S$ are coplanar. When applying the KVCM we expect that the diffraction angle $\angle(\vec{k}_R,\vec{k}_S)\geq 2\Theta_B$ for $\Theta\leq\Theta_B$ and $\angle(\vec{k}_R,\vec{k}_S)< 2\Theta_B$ else. On the contrary for the BVM there is always an {\em excess} angle $\varepsilon>0$ for any deviation from the Bragg condition as shown in Fig. \ref{fig:2ddiff}.
For the tilt geometry the KVCM again predicts that the diffracted beam lies in the $x-z$-plane, as the surface vector normal does not play a role. The BVM on the other hand implies that the diffracted beam moves out of this plane for the off-Bragg geometry as shown in Fig. \ref{fig:3dbvm}. Thus for the KVCM we arrive at the same situation as in planar geometry \ref{it:1}, see Fig. \ref{fig:2ddiff} (dashed lines). When employing the BVM instead, however, the diffracted beam is out of the $x-z$-plane upon off-Bragg replay, i.e., the spot moves mostly up/down on a screen perpendicular to the diffracted beam for in-Bragg. The corresponding situation is shown in Fig. \ref{fig:3donscreen} ({\em Left}).
The above mentioned differences as predicted by the KVCM and BVM are easy to verify in experiments and allow - next to the fact that all processes are elastic - to discriminate between the methods.
    \section{Experimental\label{sec:exp}}
\begin{figure}
\resizebox{\columnwidth}{!}{%
\includegraphics{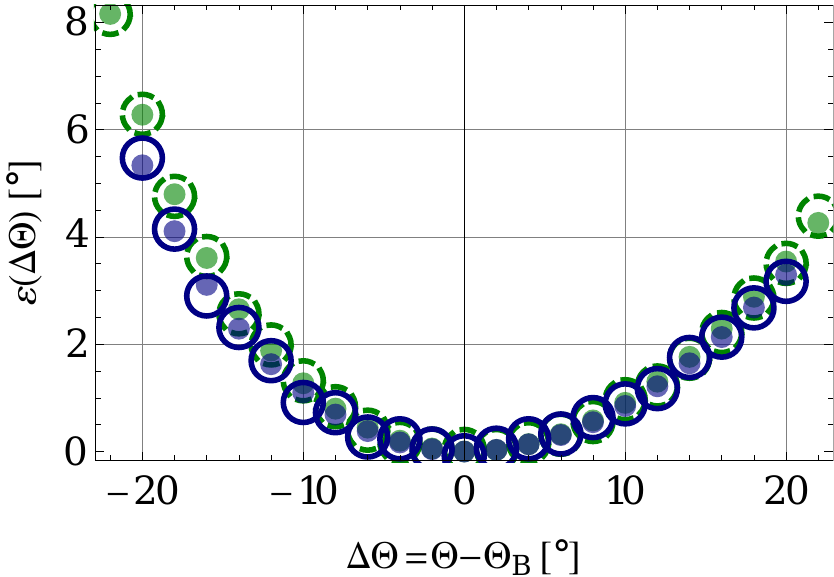}
}
\caption{Measured excess angle $\varepsilon(\Delta\Theta)$ for the experimental data at $\lambda=514$ nm (solid circle)  and $\lambda=476$ nm (dashed circle). The solid disks are the simulated values according to Eq. (\ref{eq:excess2d}). Note that the model is free of any adjustable  parameter.}
\label{fig:2dexp}       
\end{figure}

Light optical diffraction experiments were conducted on a $\rm SiO_2$ nanoparticle-polymer grating \cite{Suzuki-ao04} with a grating spacing $\Lambda=2\pi/|\vec{K}|=500$ nm using an $s$-polarized Ar-ion laser beam at a readout wavelength $\lambda=514.5$ nm unless noted otherwise. To prepare the beam properly (plane wave), we used a beam expansion system followed by a diaphragm with a diameter of about 4 mm. The effective thickness of the grating at zero tilt angle  was found to be $d_0=58.5\pm0.05 \mu$m by fitting Eq. (\ref{eq:diffeff})  to the angular dependence of the diffraction efficiency with a coupling strength of $\nu(\theta_B)\approx 1.2\pi/4$. A schematic of the setup is shown in Fig. \ref{fig:geo}. The sample is placed on an Eulerian cradle and rotated around a vertical axis for an angle $\Theta$ with or without a tilt $\Zeta$. The screen was placed at  a distance of 265 mm from the sample, so that for Bragg incidence the beam hit the screen perpendicularly. To monitor the position of the beam we 
took photos of the beam spot on the screen using a digital camera (Canon EOS D10) in auto-exposure mode. Then each photo was analyzed as follows: as a starting point we found the position of maximum intensity, followed by calculating the spot's center of gravity in a region around the maximum with about the size of the beam's diameter  (4 mm $\stackrel{\scriptscriptstyle\wedge}{=}$ 10 pixel). This was defined as the spot's position. The size of the open circles in Figs. \ref{fig:3donscreen}  and \ref{fig:2dexp} reflects the spot's size.

In planar geometry  (case \ref{it:1}, $\Zeta=0$) we observed the excess angle $\varepsilon(\Delta\Theta)>0$ upon off-Bragg replay as shown in Fig. \ref{fig:2dexp}, i.e., the spot on the screen moves in the $x-z$-plane and always in the same direction independent of the off-Bragg rotation sense. This behavior is in contradiction to the KVCM. In checking the validity of the BVM we employ Eq. (\ref{eq:excess2d}) and find that the model fits the data {\em without any adjustable parameter}. To emphasize this we performed the experiment also for another read-out wavelength $\lambda=476$ nm and the model fits as well.
While we already falsified the KVCM, the latter is a another strong indication for the applicability of the BVM instead.

Next we turn to the even more striking case of tilting the grating around the grating vector $\vec{K}$ about an angle of $\Zeta\approx45^\circ$. This results in a larger effective thickness $d$ of the grating, thus higher angular selectivity and stronger effective coupling as well. This feature is utilized as suggested in Ref.  \cite{Somenkov-ssc78} in recent diffraction experiments from holographic gratings using cold and very cold neutrons to tune the diffraction efficiency for beam splitters or mirrors  \cite{Fally-prl10,Klepp-pra11,Klepp-inprep12}. First we checked the angle of total reflection. The measured value is $(17.3\pm0.05)^\circ $, the theoretically expected one $\Theta_{TR}(\Zeta=45°)-\Theta_B=17.0^\circ$ according to Eq. \ref{eq:cutoffgen}, which fits quite well. Next we monitored the position of the diffracted beam while rotating by an angle $\Delta\Theta$ off-Bragg.
Fig. \ref{fig:3donscreen} ({\em Right}) shows the experimentally observed position of the diffraction spots on a screen upon variation of $\Delta\Theta$. The graph compiles all photographs for a rotational increment of $2^\circ$ from $\Delta\Theta=-22^\circ\ldots 8^\circ$. This graph impressively demonstrates that the direction of the diffracted beam is out of the $x-z$-plane, thus being contradictory to the KVCM. To compare the measured data with the predictions of the BVM we fitted the model to the experimental data by using a single parameter - the tilt angle $\Zeta$ - which in principle can be also just measured. Minimizing the least-squares error yielded an angle of $\Zeta=41.7^\circ$. The fitted positions are given by the solid disks, whereas the centers of gravity for each diffraction spot are indicated by dashed circles together with the photographs of the diffracted spots in the background. The agreement between the data and the positions predicted by the BVM is excellent.
    \section{Discussion}
In an excellent review {\em Russell} \cite{Russell-pr81} compared the dynamical diffraction theory,i.e, a two-eigenmodes approach, with the two-wave first order coupling theory as derived by Kogelnik and came to the conclusion that the expressions for the amplitudes are exactly analogous. Knowing this we gave a correspondence of the relevant parameters, i.e., terms for neutrons \cite{Klepp-nima11} to facilitate the communication between people working in light- or neutron optics. However, it was somewhat astonishing that the results should be exactly the same as 'In the modal method, the boundary assumes paramount importance, acting as a strong discontinuity(\ldots). In the coupled-wave picture the boundary has a very subsidiary role.'\cite{Russell-pr81}. Moreover, for the latter a class of different boundary conditions are permitted \cite{Syms-90}  which are expected to lead to different amplitudes, too.
For the KVCM the sample surface normal does not appear in the derivation at all, whereas for the BVM phase matching at the boundary is required exactly as for the DDT. So why do KVCM and DDT lead to the same expressions? In short the answer is as follows: in the derivation of the dispersion surface (DDT)  the quartic equation in the magnitude of the permitted wavevectors is reduced to a quadratic one. This approximation is valid for small deviations from the Bragg condition. The dispersion surface then consists of hyperbolic branches for the permitted wavevectors in the grating region. Without this approximation the equation to solve is in fact of 4th order and such a deviation is expected to occur for far off-Bragg replay. For this reason the approximate DDT and the KVCM lead to the identical diffraction amplitude dependencies. However, when considering the full dispersion surface we end up with a different expression that is very well consistent with the BVM. A detailed analysis will be given elsewhere.

\begin{figure}[p]
\resizebox{\columnwidth}{!}{%
\includegraphics{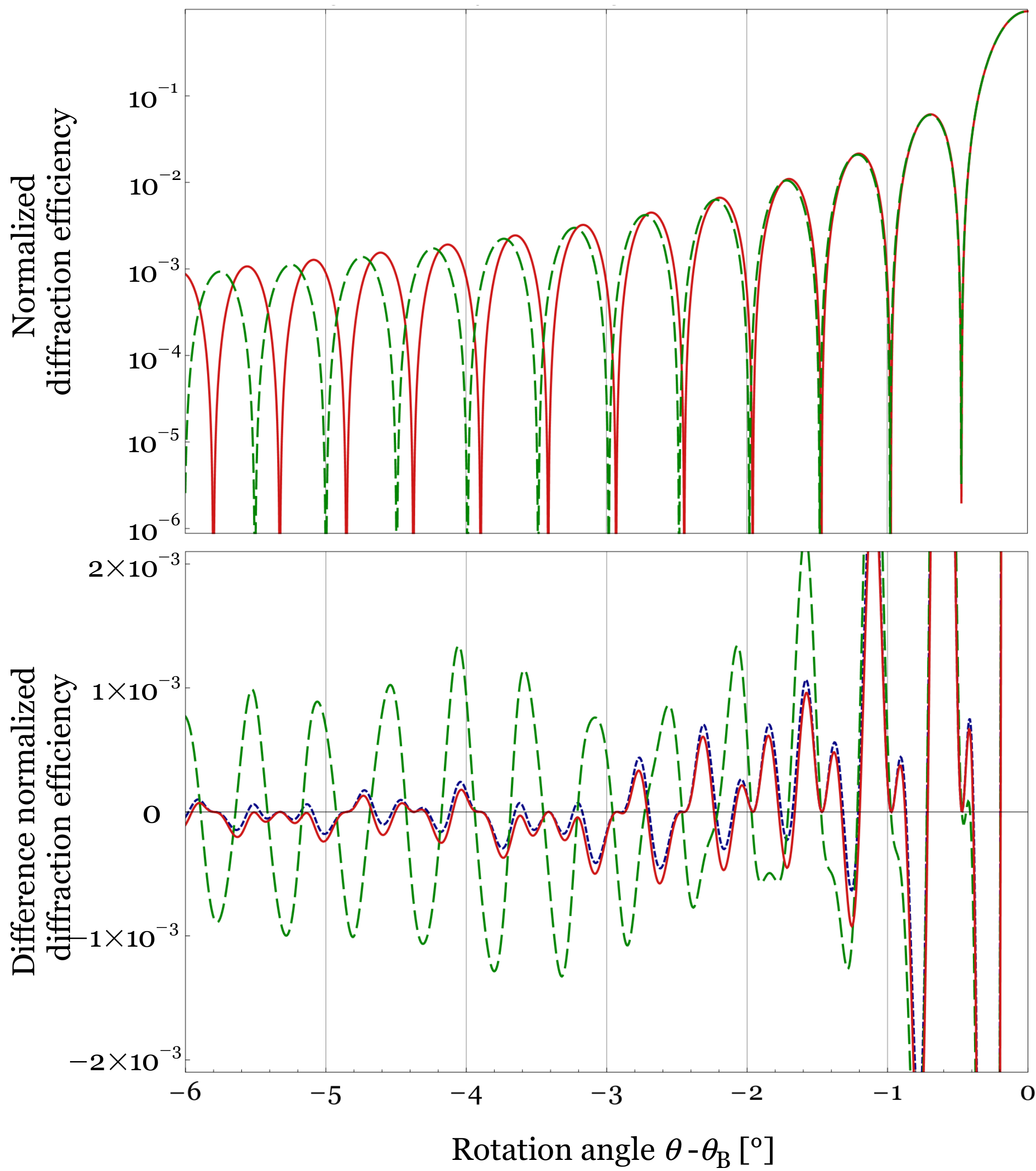}
}
\caption{{\em Top}: Comparison of the normalized first order diffraction efficiency upon off-Bragg replay for Uchida's method (BVM, red solid) and Kogelnik's method (KVCM, green dashed), cf. also Ref. \cite{Sheridan-jmo92}.
 {\em Bottom}: Difference of the normalized first order diffraction efficiency to the two-wave rigorous coupled wave a\-na\-ly\-sis upon far off-Bragg replay
 for Uchida's BVM (red solid line), Kogelnik's KVCM/approximate DDT (green dashed line), and the exact DDT without approximation (blue dotted line).}
\label{fig:comparisonDE}       
\end{figure}

The normalized diffraction efficiency, i.e., $\eta(\theta)/\eta(\theta_B)$,  upon off-Bragg replay for the BVM and the KCVM is depicted in Fig. \ref{fig:comparisonDE} ({\em Top}).
The difference between diffraction efficiency according to RCWA and the above discussed theories, i.e., Koglenik's KVCM, Uchida's BVM and the exact DDT are potted in Fig. \ref{fig:comparisonDE} ({\em Bottom}).
One can clearly notice that the BVM and the exact DDT yield very similar results, in particular when considering the zeros of the diffraction efficiency, which are determined by the boundary conditions.
When now considering the case of neutrons (x-rays) the corresponding derivations given in, e.g., Refs. \cite{Rauch-78,Sears-89,Batterman-rmp64} always treat only small deviations from the Bragg condition and thus approximate the Ewald spheres by planes. In addition, due to the fact that the refractive index is very close to unity, i.e. $1\pm 10^{-5}$, the boundary conditions have only negligible influence on the dephasing behaviour of the diffraction amplitude. In this case any of the theoretical methods describes the experimental results equally well.

We want to conclude the discussion with  a few remarks. It was noted in literature that the tilt geometry used here can be utilized to achieve polarizing effects, e.g., a polarizing beam-splitter \cite{Syms-oa85} as is obvious already by having a look on the rigorous theory \cite{Moharam-josa83.2}. In the course of the treatment given above we did not make any analysis of the diffracted beam's polarization state while keeping the polarization state of the incoming beam perpendicular to the $x-z$-plane. Yet, it is obvious that whenever the direction of the diffracted beam for this geometry is needed, reasoning on the basis of Kogelnik's KVCM fails, see e.g. \cite{Lopez-joa99}.

A second remark relates to a cutoff mentioned in Ref. \cite{Syms-ao83}. In this context we want to emphasize that the total reflection angle discussed above (see Eq. \ref{eq:cutoff2D}) is not this cutoff. The difference is as follows: in our case a diffracted beam is excited in the medium but cannot be coupled out via a parallel grating slab. In contrast the cutoff angle defines the critical value for which a (higher) diffraction order cannot be excited any more (evanescent wave).

The results for the direction of the diffracted beam in off-Bragg replay might be important for the implementation of a novel beam coupling technique \cite{Fritzenwanker-10}. The latter is an improved variant of the grating translation technique to determine the amplitudes and possible phases of mixed refractive index/absorption gratings \cite{Sutter-josab90,Kahmann-josaa93,Fally-josaa06}.

Finally, it is interesting to note that the kinematical theory of diffraction - based on the first Born approximation and hence neglecting the effect of multiple scattering -  already yields the correct direction of the diffracted beam, see e.g. \cite{Sears-89}, and in this respect the KVCM was a step backwards. We thus recommend using the BVM because it is as easy to implement as the KVCM and gives overall correct results.

Let us pose a final question that might have been already anticipated by the reader:  Why did this  - now obvious - discrepancy between theory and experiment not limit the popularity
of the corresponding paper?
One argument could be that people just employ the theory to determine the relevant materials parameters. In this case typically far off-Bragg replay does not play a role. Another could be the fact that {\em Uchida} in Ref. \cite{Uchida-josa73} focusses on attenuation of the modulation along the thickness rather than on the different choice of wavevector mismatch ('no essential difference').
As noted in Ref. \cite{Syms-ao83}, the differences between the exact and the approximate theory will be recognized only for 'large deviations from the Bragg condition, namely, in the sidelobe structure for comparatively thin holograms'. As our nanoparticle-polymer composites \cite{Suzuki-ao04} have a large coupling strength one can reach considerable diffraction efficiencies already at moderate thickness in the range of 10 $\mu$m as opposed to the extensively investigated photorefractive electrooptic crystals \cite{Solymar-02,Yeh-93}.
    \section{Conclusion}
It is known for decades that the amplitude of a beam diffracted by a volume grating is usually well described by the coupled wave theory using the KVCM (Kogelnik). The less prevalent BVM (Uchida) was claimed to be superior for off-Bragg replay but the diffraction efficiency does not allow to discriminate easily between the two methods \cite{Sheridan-jmo92}. Here, we investigated the {\em direction} of the diffracted beam which impressively follows the behavior derived from the BVM method - even without any free fitting parameter -  and disagrees with the KVCM.

\noindent{\bf Acknowledgments} Financial support by the Austrian Science Fund (FWF): P-20265 is acknowledged.
We are grateful to Romano A. Rupp for drawing our attention to a mistake in Eq. (\ref{eq:excessKVCM}) and Eq. (\ref{eq:excess2d}) published in Appl. Phys. B.

\end{document}